# THE SHIFTING IMPACT OF RECURRENT FLOODING ON TRANSPORTATION ACCESSIBILITY: A CASE STUDY OF AFFECTED POPULATIONS IN THE HAMPTON ROADS REGION

**Luwei Zeng, Ph.D.**
Graduate Research Assistant
Department of Systems and Information Engineering
University of Virginia, Charlottesville, VA, 22903
lz6ct@virginia.edu

**T. Donna Chen, PE, Ph.D.**
Corresponding Author
Associate Professor
Department of Civil & Environmental Engineering
University of Virginia, Charlottesville, VA 22903
tdchen@virginia.edu

**John S. Miller, PE, Ph.D.**
Associate Director for Environment, Planning and Economics
Virginia Transportation Research Council
Charlottesville, VA, 22903
John.Miller@VDOT.Virginia.gov

**Faria Tuz Zahura, Ph.D.**
Graduate Research Assistant
Department of Engineering Systems and Environment
University of Virginia, Charlottesville, VA 22903
fz7xb@vriginia.edu

**Jonathan L. Goodall, Ph.D., PE**
Professor
Department of Civil & Environmental Engineering
University of Virginia, Charlottesville, VA 22903
goodall@virginia.edu

**Abstract**

Accelerated sea level rise has resulted in recurrent flooding in coastal regions, increasingly impacting both transportation systems and local populations. Using the Hampton Roads region in Virginia as a case study, this study a) identifies "hotspots" with frequent, significant accessibility reduction for work and non-work travel utilizing crowdsourced WAZE flood report data during the month of August over a 5-year period (2018-2022); and b) examines the shifts in social vulnerability in populations residing in these hotspots over the 5 year period using 2016 and 2021 American Community Survey data. Results show that approximately 12% and 3% of the region's population reside in hotspots experiencing significant recurrent flooding-induced accessibility reduction for work and non-work trips. Social vulnerability analysis revealed that populations with greater socio-economic and transportation vulnerabilities are more susceptible to recurrent flooding-induced accessibility impacts in terms of both extent and frequency. Furthermore, a comparison of social vulnerability indices between 2016 and 2021 shows an increasing trend of social vulnerability for highly impacted zones, with low-income, disabled, and households with young children having restricted ability to relocate from these zones. The findings reinforce the necessity for spatially and temporally disaggregated studies of climate event impacts. Furthermore, the longer-term population trends highlight the importance of dynamic assessment of climate event impacts at different time scales.

## 1. INTRODUCTION

Coastal residents are experiencing increased recurrent flooding impacts on roadways due to climate change, affecting network accessibility through travel delays. Zeng et al. (2023) developed a framework evaluating these impacts in a spatially and temporally disaggregated manner, utilizing crowdsourced flood incident reports and congestion data (*1*). This framework identifies accessibility reduction, for both work-related and nonwork-related trips, during different time periods of the day, and assesses the social vulnerability of affected populations. A case study of the Hampton Roads region, Virginia, in August 2018 demonstrated that recurrent flooding effects are unevenly distributed across the region, with locations experiencing higher accessibility reduction showing higher social vulnerability traits.

However, Zeng et al. (2023) examined the accessibility impacts of recurrent flooding at one snapshot in time (August 2018), and did not capture how the flooding impacts and the populations affected are changing over time (*1*). In this study, using the same impact assessment framework, the relationship between recurrent flooding-induced accessibility reduction and social vulnerability are examined across five years (from 2018 to 2022) using American Community Survey 5-year estimate data and crowdsourced WAZE flood incident report data. The research objectives are to (1) improve the framework developed in Zeng et al. (2023) to process crowdsourced WAZE data more efficiently, (2) identify the most frequently impacted locations of recurrent flooding over the five-year time span, and (3) analyze any changes in social vulnerability of the highly impacted areas between 2016 and 2021. This study again focuses on the Hampton Roads region, Virginia—a region of roughly 3,700 mi$^2$ with 1.7 million people and 1 million jobs.

## 2. LITERATURE REVIEW

Methods used in the existing literature for quantifying the impacts on flooding include numerical models, simulation and prediction models (*2–9*). A significant body of previous literature focused on the impact of major climate events on transportation accessibility (*5, 10–12*), and a smaller body of studies explored the effect of recurrent flooding events (*1, 3, 13*). The social vulnerability index framework is developed based on a series of studies that investigate the relationship between social vulnerability and transportation accessibility under various climate event contexts (*1, 14–17*). For this study, the literature review focuses on studies that examine the impacts of climate events on specific populations, especially those that have examined these impacts over time.

A considerable amount of literature has considered the impact of climate events and natural disasters on the population. These studies show that due to various exposures and social vulnerability characteristics, such impact is disproportionally distributed across regions and individuals, and low-income populations are more vulnerable than high-income populations as they suffer higher direct damages (*18–23*). These studies emphasize the role of socioeconomic status in disaster resilience. In the transportation context, Freitas et al. (2019) found that mobility is worse for vulnerable populations living in areas most susceptible to impacts of climate change, and adding mobility analysis into social vulnerability calculations will lead to an increase in the total vulnerability (*17*). Zandt et al. (2012) found that individuals with limited access to private transportation exhibited decreased rates of evacuation during disaster events. These studies all focused on large-scale natural disasters and ignored the impacts of recurrent smaller scale events. The cumulative impacts of recurrent smaller events are compounding over time, as recurrent flooding events become more frequent and grow in intensity (*24*).

Several studies have investigated the shift in social vulnerability characteristics due to natural disasters. Donner & Rodríguez (2008) assessed changes in socially vulnerable groups in the U.S from 1980 to 2000, and emphasized the need for frequent updating of population characteristics in disaster mitigation (*23*). Cutter and Finch (2008) explored the temporal and spatial variations in social vulnerability under natural hazards using county-level census data from 1960 to 2000, finding a general reduction but more scattered distribution of vulnerability over time (*25*). Zhou et al. (2014) analyzed trends at the county-level in China from 1980 to 2010, identifying a pattern of clustering, scattering, and re-clustering of socially vulnerable groups (*26*). These studies underscore the necessity for disaggregated regional and local analysis of social vulnerability under climate events, highlighting the increasing frequency and severity of such events exacerbate the vulnerability of populations in hazardous areas, and emphasizing the importance of differentiating between local and national impacts.

Research examining the long-term shifts in social vulnerability to transportation under climate disruption is scant. Gray and Mueller (2012) presented a 15-year period case study in rural Bangladesh, revealing that disaster flooding unevenly impacts mobility, with women and the poor population being more vulnerable (*27*).

In summary, existing literature shows that natural disasters and climate events disproportionately affect socially vulnerable populations, with poverty as a consistent key factor. Existing literature highlights that social vulnerability is dynamic, and there is a shortage of localized time-space examination. This research gap led to the current study's focus on small-scale climate events, specifically recurrent flooding, and its multi-year impacts on transportation accessibility. The study also uses census data to analyze the changing social vulnerability of the population living in zones highly impacted by recurrent flooding.

## 3. DATA SOURCES AND PRE-PROCESSING

The analysis of accessibility and socially vulnerable characteristics integrate various data sets, which includes the local roadway network, employment destinations, the non-work destinations, roadway flooding incident reports and subsequent traffic impacts, and the socioeconomic characteristics of the residents in each analysis zone. Maps throughout this paper were created using ArcGIS® software by Esri.

### 3.1 Transportation Network Data and GIS Layer

The GIS shapefiles for the study area are provided by the Hampton Roads Transportation Planning Organization (HRTPO) and contain the population and employment characteristics for each of the 1,173 traffic analysis zones (TAZs) in the region. The shapefile for the roadway network was acquired from the Virginia Geographic Information Network (VGIN). The file includes details such as road names, directions, local speeds, and the number of lanes (*28*). This information allows for the construction of a network dataset using ArcGIS Pro software and the calculation of travel impedance to evaluate accessibility metrics.

### 3.2 Crowdsourced Points-of-Interest (POI) data

OpenStreetMap (OSM), which is free and available to the public, provided the location of non-work, major points-of-interest (POI). In this study, the major POI within the Hampton Roads region includes healthcare, public service, entertainment, food and drink, education, and banks. A total of 2,078 POIs were extracted from the OSM applications programming interface (API) Overpass using Python. Table 1 shows the number of POIs in each category.

**Table 1 POIs in the Hampton Roads Region**

| Healthcare | Public Service | Entertainment | Food & Drink | Education | Banks | Total |
|---|---|---|---|---|---|---|
| 152 | 190 | 486 | 960 | 232 | 58 | 2078 |

### 3.2 Socioeconomic Data

The GIS layers from HRTPO contain 2015 data on population and employment for each TAZ. To analyze changes in socioeconomic characteristics in highly impacted zones and to compute the social vulnerability index across five years, the American Community Survey (ACS) 5-year estimates for 2016 and 2021 at the census tract level were used. The 5-year estimate data offers detailed socioeconomic information relevant to social vulnerability. The use of 2016 and 2021 data prevents overlapping in the 5-year collection process. Lastly, as TAZs and census tracts have mismatched boundaries, each TAZ was linked to the census tract that contains the largest portion of its area.

**3.3 Crowdsourced Flood Incident and Traffic Data**

Mobile navigation application WAZE collects user-provided flood incident reports, and associated traffic condition data which is recorded passively. The WAZE Cities data-sharing program makes the aggregated user-reported incident data available to researchers. Based on WAZE data from 2018 to 2020, Figure 1 shows that the month of August typically has the highest number of flood incident reports out of the year, suggesting that August is typically associated with significant flooding situations. Thus, in this study, flood incident reports and traffic data for the month of August across five years, from 2018 to 2022, are used for analysis.

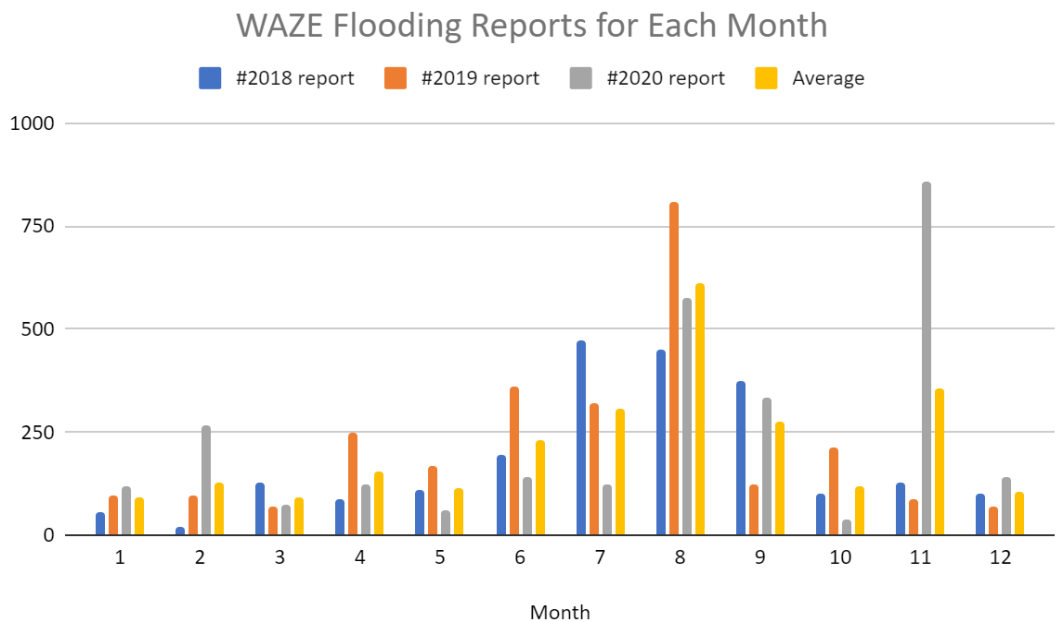

**Figure 1 WAZE flooding reports for each month[1]**

Both the flood incident reports and the traffic data are time-stamped and location-specific. To mitigate the effect of error or misinformation that is inherent within crowdsourced data (29), the flood incident reports were verified with traffic condition data. A 50-meter buffer is applied to the location of each flood incident report. If any congestion data is recorded within the buffer area by WAZE traffic data, then this flood report is redeemed verified, and the flood incident report location is assigned the associated congestion speed. If no congestion data is recorded within the buffer area, then the flood report is considered of no (or low) impact to transportation accessibility and is excluded from analysis.

The flood reports for the month of August from 2018 to 2022 were aggregated across five time-of-day periods: period 1 (12:00 to 6:00 am), period 2 (6:00 to 9:00 am), period 3 (9:00 am to 3:00 pm), period 4 (3:00 to 6:00 pm), and period 5 (6:00 pm to 12:00 am), matching the time-of-day disaggregation in the regional travel demand model. The time-period specific aggregation process is done by building a customized GIS toolbox using the model builder function in ArcGIS Pro 3.0, described in detail in the Methodology section of this paper. Table 2 shows the summary of flood incident reports across five years before and after the verification with associated congestion data.

---

[1] 2021 and 2022 data omitted from Figure 2 due to incomplete data for some months.

**Table 2 Flood incident reports per time-of-day period**

| Time Period | # Original reports | # Reports associated with congestion data | % Reports associated with congestion data |
|---|---|---|---|
| Period 1 (12:00am to 6:00am) | 3167 | 1752 | 55.32% |
| Period 2 (6:00am to 9:00am) | 775 | 270 | 34.84% |
| Period 3 (9:00am to 3:00pm) | 4126 | 2275 | 55.14% |
| Period 4 (3:00pm to 6:00pm) | 2182 | 1184 | 54.26% |
| Period 5 (6:00pm to 12:00am) | 3781 | 2075 | 54.88% |

## 4. METHODOLOGY

This study evaluated work-related accessibility based on employment locations and non-work-related accessibility based on POIs, using the TAZ as the spatial unit of analysis. Of the 1,173 TAZs, 1,115 with non-zero population were considered as origins, and 990 and 670 TAZs with non-zero employment and POIs, respectively, as destinations. Each origin was paired with each destination to create the Origin-Destination (OD) matrix. Travel time was calculated using driving time between TAZ centroids with non-zero population, jobs, or non-work POIs. Subsequent sections detail the integration of WAZE congestion data, the generation of OD matrices to compute flood-induced accessibility changes, and the social vulnerability index (SVI) framework for assessing changes in the most affected zones.

**4.1 Development of WAZE data processing toolbox**

The conversion of WAZE congestion data into line barriers on the base roadway network in GIS in order to generate the travel cost under flooding conditions requires realigning the congestion data to the roadway network (via projection and data cleaning), as they are not a perfect match. To increase efficiency, four customized tools were built using model builder in ArcGIS Pro to automate the process. The key process of this workflow is to produce a line barrier layer by clipping the base roadway network segment (that can represent the closest congestion lines), and spatially joining the congestion data on to it. Table 3 describes the function of each customized tool in detail.

**Table 3 Description of tools in customized toolbox**

| Tool | Function |
|---|---|
| Tool 1: Process WAZE data: flooding report locations & congestion lines | • Convert locations of flood reports into points layer and congestion data into polylines layer.<br>• Apply 50-meter radius buffer for each flood report point, filtering for verified flood reports that have associated congestion data.<br>• Tally the number of original reports and filtered reports.<br>• Select the valid flooding reports and congestion lines.<br>• Overlay the base roadway network with the congestion lines and produce a new polyline layer, which is the collection of segments of the base roadway network that is closest to congestion lines, named "draft line barriers". Create a buffer with 1.5-meter-wide polygon layer |

| | | around the line barriers called "buffer on draft line barriers". |
|---|---|---|
| Tool 2A: Cleaning congestion lines | | • Create points at segment intersections.<br>• Apply 6.5-meter buffers at intersection points, named "intersection area".<br>• Use Erase tool for "draft line barriers" by "intersection area" (both are polygon layers). Thus the "draft line barriers" which overlap with "intersection areas" are deleted.<br>• Select and remove the segments that are not within the updated "draft line barriers" using Select by location tool.<br>• Produce a "clean line barriers" polyline layer that is spatially matched to the base roadway network. |
| Tool 2B: Dissolve overlapping congestion lines | | • Merge multiple congestion lines at the same location into one line to form polyline layer "clean congestion lines."<br>• Calculate the average congested driving speed for each merged congestion line. |
| Tool 3: Spatially join congestion lines and line barriers | | • Spatially join layers "clean congestion lines" with "clean line barriers" with the spatial relationship of "closest".<br>• Calculate the scaling factor using $\frac{Local\ Speed}{Congestion\ Speed}$.<br>• Produce a "final line barriers" polyline layer with scaling factor. |

This set of automated processes significantly reduces the need for manual intervention in data processing. With this toolbox, the inputs are WAZE flood incident reports and congestion data, and the outputs are the number of original and verified flood incident reports, and the congestion line barriers layer with speed scaling factor. This line barriers layer can be used directly in the OD matrix generation for accessibility calculations. However, the congestion data projection still requires some manual inspection to ensure that extra segments are removed at the roadway intersection from the line barriers conversion (Tool 2A).

**4.2 Accessibility calculations**
The accessibility calculation uses Equation 1 shown below. The impedance (friction factor) of travel from $i$ to $j$, $C_{(ij)}$, is a function of travel time $t_{ij}$, calculated by Equation 2.

$$A_i = P_i \sum_j O_j f(C_{ij}) \tag{1}$$

Where:
    $A_i$: access from TAZ $i$.
    $P_i$: population of TAZ $i$.
    $O_j$: number of opportunities available at destination TAZ $j$.
    $C_{ij}$: cost of travel from TAZ $i$ to $j$.
    $f(C_{ij})$: impedance function, also known as friction factor.

$$F_{ij}^p = a * t_{ij}^b * exp(c * t_{ij}) \tag{2}$$

Where:
    $F_{ij}^p$: friction factor for trip purpose $p$ from TAZ $i$ to $j$.

$t_{ij}$: the travel impedance between TAZs $i$ and $j$, travel time.
$a$ : gamma function scaling factors, does not change the shape of the function.
$b$ : gamma function scaling factors, always negative.
$c$ : gamma function scaling factors, generally negative.

For the Hampton Roads region, the chosen parameters in Equation 2 are gamma-based impedance scaling factors sourced from a database of practice from major metropolitan planning organizations (MPOs) (*30*), selected to align with the region's long-term planning process. The long-range regional model incorporates a gravity-based metric to represent travel impedance. In terms of weighing travel time, parameters in line with values for "MPO 1" were chosen due to their moderate sensitivity concerning work-related trips (Cambridge Systematics et al., 2012), where *a=2*, the values of parameters for home-based work trips are b = -0.503 and c = -0.078, and for home-based non-work trips, the parameters are b = -3.993 and c = -0.019.

Travel cost for each OD pair $t_{ij}$ was generated using ArcGIS Pro 3.0's network analysis function, based on roadway shapefiles from HRTPO. A baseline travel time was established using the free-flow speed traffic network, serving as comparison for flood scenarios. Within the ArcGIS Pro network analysis, travel time for each OD pair is stored in the line layer's attribute table. A layer representing restrictions or reductions in access, described as "final line barriers", is provided in the OD matrix generation tool, simulating temporary network changes. This layer's cost is scaled to represent travel costs when barriers are traversed (Tool 3). The original OD matrix's travel cost is updated with line barriers, reflecting flood-related delays and used in flood scenario accessibility calculations. By adjusting origin and destination input, the process analyzes accessibility change for both work and non-work trips. Calculations are performed in R studio using Equations (1) and (2).

The percentage change in accessibility from the base case to the flood disruption scenario is computed based on Equation (3) for each origin TAZ.

$$\Delta A_i = \frac{A_{b,i} - A_{f,i}}{A_{b,i}} * 100 \qquad (3)$$

Where:
$\Delta A_i$: accessibility difference between base scenario and flood scenario of TAZ *i*.
$A_{b,i}$: accessibility for TAZ *i* in base scenario.
$A_{f,i}$: accessibility for TAZ *i* in flood disruption scenario.

Based on Equation (3), the larger the value of $\Delta A_i$, the greater the impact of recurrent flooding on accessibility in TAZ *i*. Note that because the population *$P_i$* of a given zone does not change from the base scenario to the flood scenario, the term *$P_i$* in Equation (1) does not affect the results in Equation (3).

**4.3 Social vulnerability index**
The SVI framework illustrated in Figure 2 is implemented for exploring the characteristics of communities whose accessibility is highly affected by recurrent flooding. This framework is based on CDC methodology (*31*) and follows the structure provided by Masterson et al. (*32*).

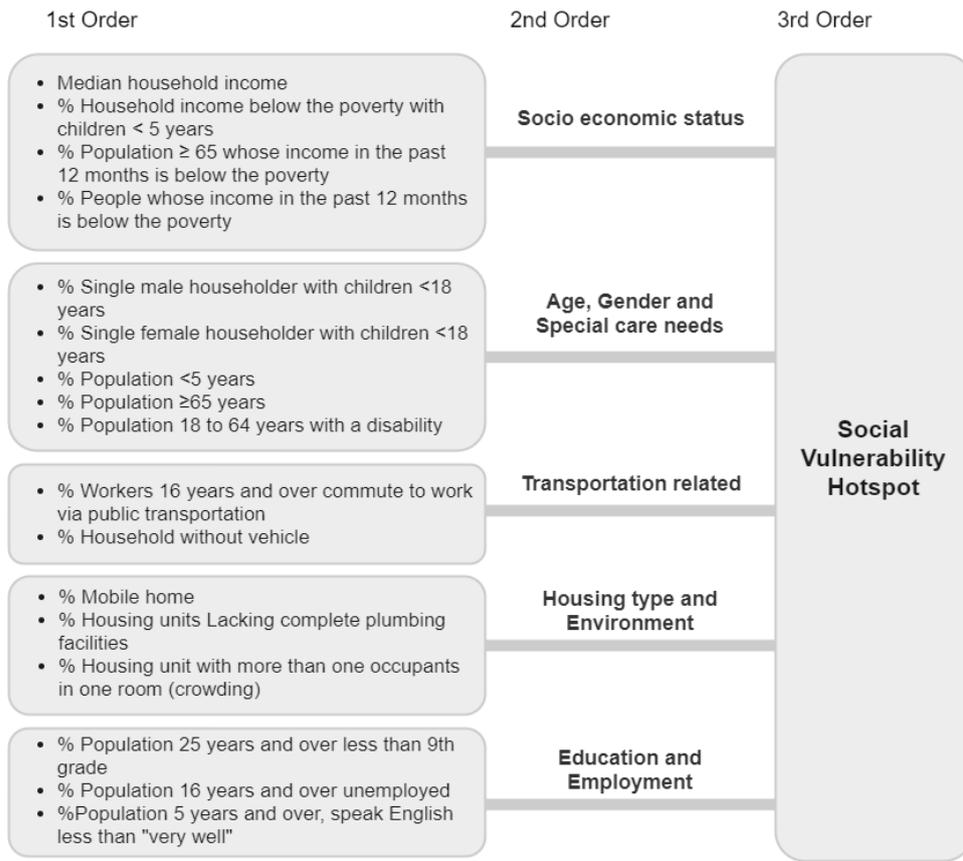

**Figure 2 Social vulnerability index (SVI) framework**

The SVI first order variables can be found in the ACS 5-year estimate data. The values for each TAZ are then normalized (across all TAZs) using the min-max scaling method shown in Equation 4.

$$x_{scaled} = \frac{(x - x_{min})}{(x_{max} - x_{min})} \qquad (4)$$

Where:
$x_{scaled}$ : scaled value
$x$: original cell value
$x_{min}$: minimum value of the first order variable
$x_{max}$: maximum value of the first order variable

The normalized first order SVI values range from 0 to 1, with higher values indicating greater vulnerability. The second order SVI is then computed by adding the normalized value of the first order SVI within each of the domains. The min-max normalization is again applied on the second order SVI values. A similar process is applied to sum second order SVI values to generate the final set of third order SVI values.

## 5. RESULTS AND DISCUSSION

Within the Hampton Roads regions, 58 TAZs have zero population and 2 TAZs are missing valid road network information in GIS, therefore the accessibility change due to recurrent flooding are assessed in 1,113 of 1,173 total TAZs, for the month of August from 2018 to 2022. TAZs with less than 1% reduction in accessibility are considered low impact. TAZs with reduction in accessibility between 1 and 5% are considered medium impact. Lastly, TAZs with reduction in accessibility greater than 5% are considered high impact. In this study, we focus on the cumulative frequency (i.e., how many times a single TAZ experiences more than 5% accessibility reduction due to recurrent flooding within the five-August study period). The results are presented in the heat maps in Figure 3(a) and 5(b) for work and non-work trips, respectively.

**5.1 Distribution of Highly Impacted TAZs**

As shown in 3 (a), for work trips, almost all TAZs in the study area experience high accessibility reductions due to recurrent flooding for at least one time-of-day period during the month of August between 2018 to 2022. Areas in cities of Norfolk, Hampton, York, James City, and a portion of Virginia Beach contain the most frequently highly affected (high accessibility reduction for 15 or more time-of-day periods) 154 unique TAZs, which contain 11.95% of the total Hampton Roads region population. As shown in 3 (b), the most affected TAZs for non-work trips are mostly concentrated in the city of Norfolk, with 31 unique TAZs containing 3.09% of the region's total population. Due to the location distribution of POIs, the accessibility impacts for work-related trips are more widespread, while such impacts on non-work-related trips are more clustered. TAZs that experience 15 or more time-of-day periods (approximately 72 hours across five Augusts) of high accessibility reduction are further analyzed in this study. We define these zones as "work critical zones" (WCZ) for work-related trips and "non-work critical zones" (NCZ) for nonwork-related trips.

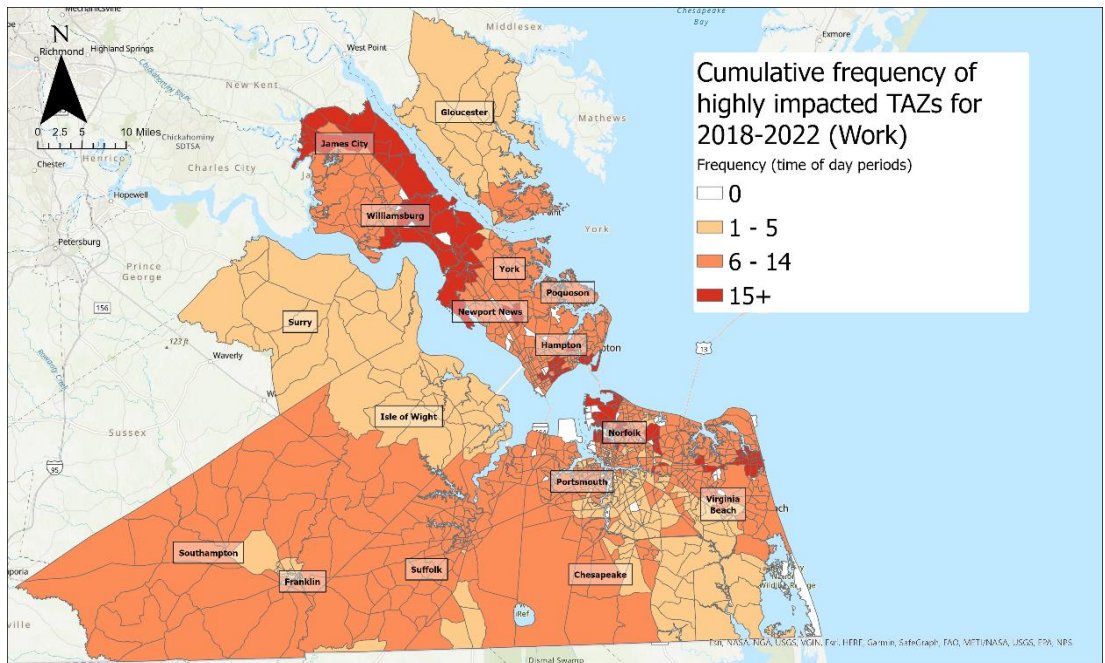

**(a)**

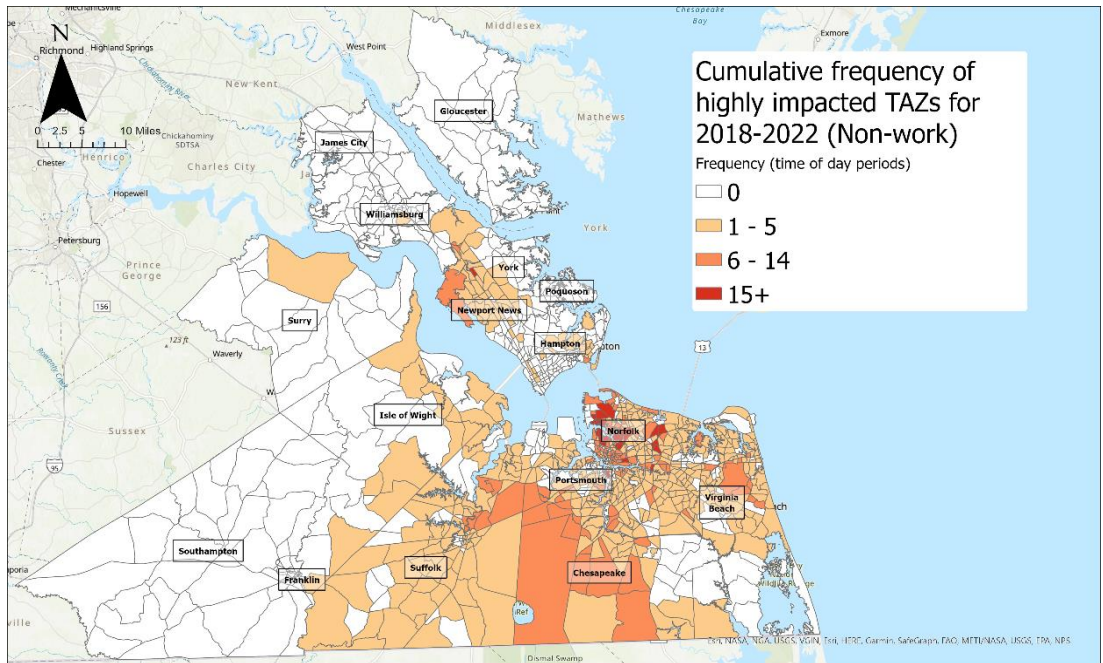
(b)

**Figure 3 Frequency of high accessibility reduction due to recurrent flooding, August 2018-2022, for (a) work, (b) non-work trips**

**5.2 Shifts in Affected Populations in Critical Zones**

Both WCZs and NCZs are further analyzed to understand the social vulnerability characteristics of these zones and how such characteristics are changing over time. Principal component analysis (PCA) is applied to the second order SVI for these selected TAZs for both trip purposes, where the second order SVIs were computed using 2016 and 2021 ACS 5-year estimate data. The relationship between second order SVI variables (socioeconomic status; age, gender, and special care needs; transportation; housing type and environment; and education and employment) and the frequency of each TAZ being impacted by recurrent flooding with high accessibility reduction (more than 5%) are examined. For 2016 ACS data, PCA results show that for work trips, only the first two principal components (PCs) have eigenvalue greater than 1, and accumulated amount of explained variances is 64.8%. Similarly, for non-work trips, the first two PCs have eigenvalue greater than 1 with 67.5% accumulated explained variance. As shown in Figure 4(a) and 4 (b), only the first two PCs were selected and presented in the biplot for work and non-work trips. For the 2021 ACS data, likewise, for both work and non-work trips, only the first two PCs have eigenvalue greater than 1, with 62.2% and 64.4% of accumulated explained variance. Figures 5 (a) and 5 (b) show the biplot for the first two PCs for work and non-work trips for 2021 ACS data.

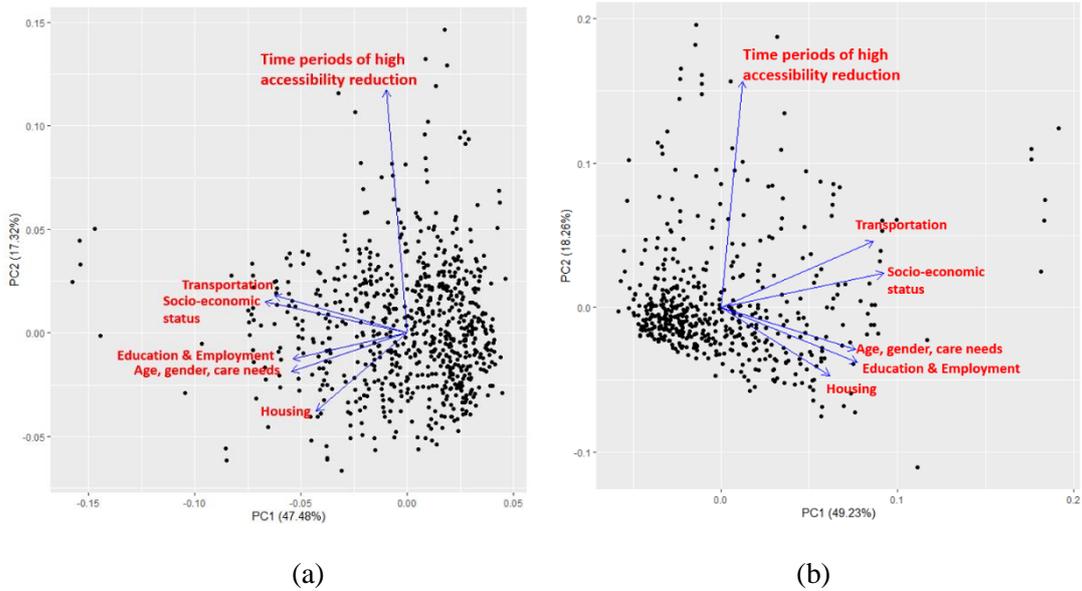

(a)                           (b)
**Figure 4 PCA biplot for 2016 ACS data (a) work trips, (b) non-work trips**

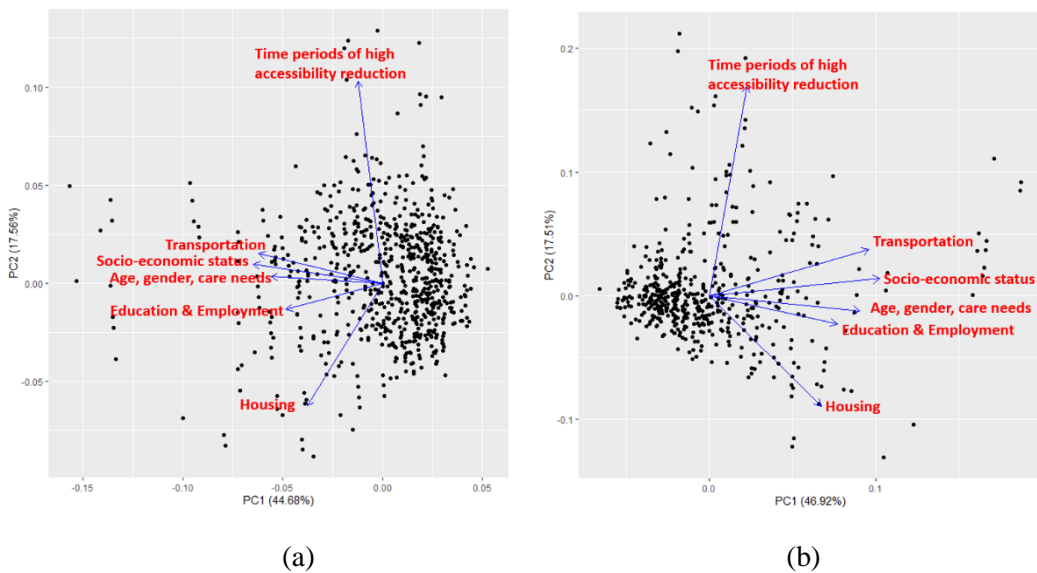

(a)                           (b)
**Figure 5 PCA biplot for 2021 ACS data (a) work trips, (b) non-work trips**

As shown in Figure 4 and 5, for both 2016 and 2021 and for both work and non-work trips, the biplots indicate that population with higher vulnerability in transportation and socio-economic status are more likely to experience higher frequency of high accessibility reduction due to recurrent flooding. This finding is consistent with that of Hallegatte et al. (2020) and Masozera et al. (2007), with both studies suggesting that poverty is a key factor for vulnerability to natural disasters. Low-income households and communities are often more adversely affected by natural disasters because they have fewer housing alternatives and tend to live in areas that are more prone to (*33*, *34*). Housing prices are relatively lower in areas with a higher risk of flood hazards compared to other locations (*35*). Other studies have indicated that smaller scale climate events such as recurrent flooding also brings considerable effects on low-income populations (*33*, *36*). Masozera et al. (2007) also found that mobility or transportation availability is an important factor in the recovery stage from a weather event, further complicating recurrent flooding's impacts on populations vulnerable in transportation access. Other second order SVI variables (age, gender, care needs, education, employment, and housing) show a negative correlation with the frequency of high accessibility reduction. To further analyze the trends of social vulnerability changes between 2016 and 2021 for critical TAZs experiencing high accessibility reduction, Table 4 provides a comparison for the input values into the SVI framework obtained from the ACS data. Figure 6 shows the percentage change in these social vulnerability indicators from 2016 to 2021, for both WCZs and NCZs.

**Table 4 Comparison of TAZ social vulnerability for 2016 and 2021**

| | | Index | 2016 (mean) | | | 2021 (mean) | | | % change 2016 to 2021 | | |
|---|---|---|---|---|---|---|---|---|---|---|---|
| | | | Hampton Roads | WCZ | NCZ | Hampton Roads | WCZ | NCZ | Hampton Roads | WCZ | NCZ |
| Population | | Population based on census data | 1,702,029 | 286848 (16.85%) | 91723 (5.39%) | 1,737,368 | 260837 (15.01%) | 74268 (4.27%) | 2.1% | -9.1% | -19.0% |
| Socio economic status | 1 | Median household income | $ 61,133 | $ 55,054 | $ 54,888 | $ 75,042 | $ 68,899 | $ 66,668 | 22.8% | 25.1% | 21.5% |
| | 2 | % Household income below the poverty with children < 5 years | 15.15 | 17.83 | 13.71 | 11.24 | 14.18 | 11.01 | -25.8% | -20.5% | -19.7% |
| | 3 | % Population ≥ 65 whose income in the past 12 months is below the poverty | 8.41 | 9.76 | 10.41 | 9.00 | 9.78 | 9.82 | 7.0% | 0.3% | -5.6% |
| | 4 | % People whose income in the past 12 months is below the poverty | 13.93 | 17.04 | 17.30 | 11.85 | 14.88 | 15.43 | -14.9% | -12.7% | -10.8% |
| Care needs | 5 | % Single male householder with children <18 years | 2.21 | 2.27 | 2.19 | 1.35 | 1.29 | 1.41 | -38.7% | -43.0% | -35.6% |
| | 6 | % Single female householder with children <18 years | 9.39 | 9.01 | 9.27 | 7.24 | 6.59 | 6.57 | -22.9% | -26.8% | -29.2% |
| | 7 | % Population <5 years | 6.35 | 5.88 | 5.95 | 6.30 | 6.71 | 7.10 | -0.9% | 14.1% | 19.4% |
| | 8 | % Population ≥65 years | 10.16 | 10.07 | 9.56 | 15.39 | 14.95 | 12.93 | 51.4% | 48.5% | 35.3% |
| | 9 | % Population 18 to 64 years with a disability | 13.42 | 12.73 | 11.12 | 11.45 | 12.06 | 11.08 | -14.7% | -5.3% | -0.3% |
| Transportation | 10 | % Workers 16 years and over commute to work via public transportation | 2.33 | 2.09 | 3.94 | 1.69 | 2.32 | 2.13 | -27.6% | 11.1% | -46.0% |
| | 11 | % Household without vehicle | 7.90 | 8.42 | 9.88 | 7.34 | 9.00 | 8.66 | -7.1% | 6.8% | -12.3% |
| Housing | 12 | % Mobile home | 2.21 | 1.61 | 0.71 | 2.09 | 1.92 | 1.10 | -5.4% | 18.9% | 55.3% |
| | 13 | % Housing units lacking complete plumbing facilities | 0.24 | 0.24 | 0.13 | 0.43 | 0.45 | 0.26 | 76.6% | 88.9% | 92.4% |
| | 14 | % Housing unit with more than one occupants in one room (crowding) | 2.01 | 2.63 | 2.44 | 1.65 | 1.56 | 1.65 | -17.8% | -40.6% | -32.5% |
| Education & Employment | 15 | % Population 25 years and over less than 9th grade | 3.24 | 3.42 | 2.74 | 2.55 | 2.65 | 2.93 | -21.2% | -22.8% | 7.0% |
| | 16 | % Population 16 years and over unemployed | 4.90 | 5.17 | 4.85 | 3.41 | 3.61 | 4.65 | -30.5% | -30.1% | -4.1% |
| | 17 | %Population 5 years and over, speak English less than "very well" | 2.80 | 3.76 | 2.87 | 2.79 | 3.49 | 3.23 | -0.3% | -7.1% | 12.7% |

**Figure 6 Change in social vulnerability indicator from 2016 to 2021**

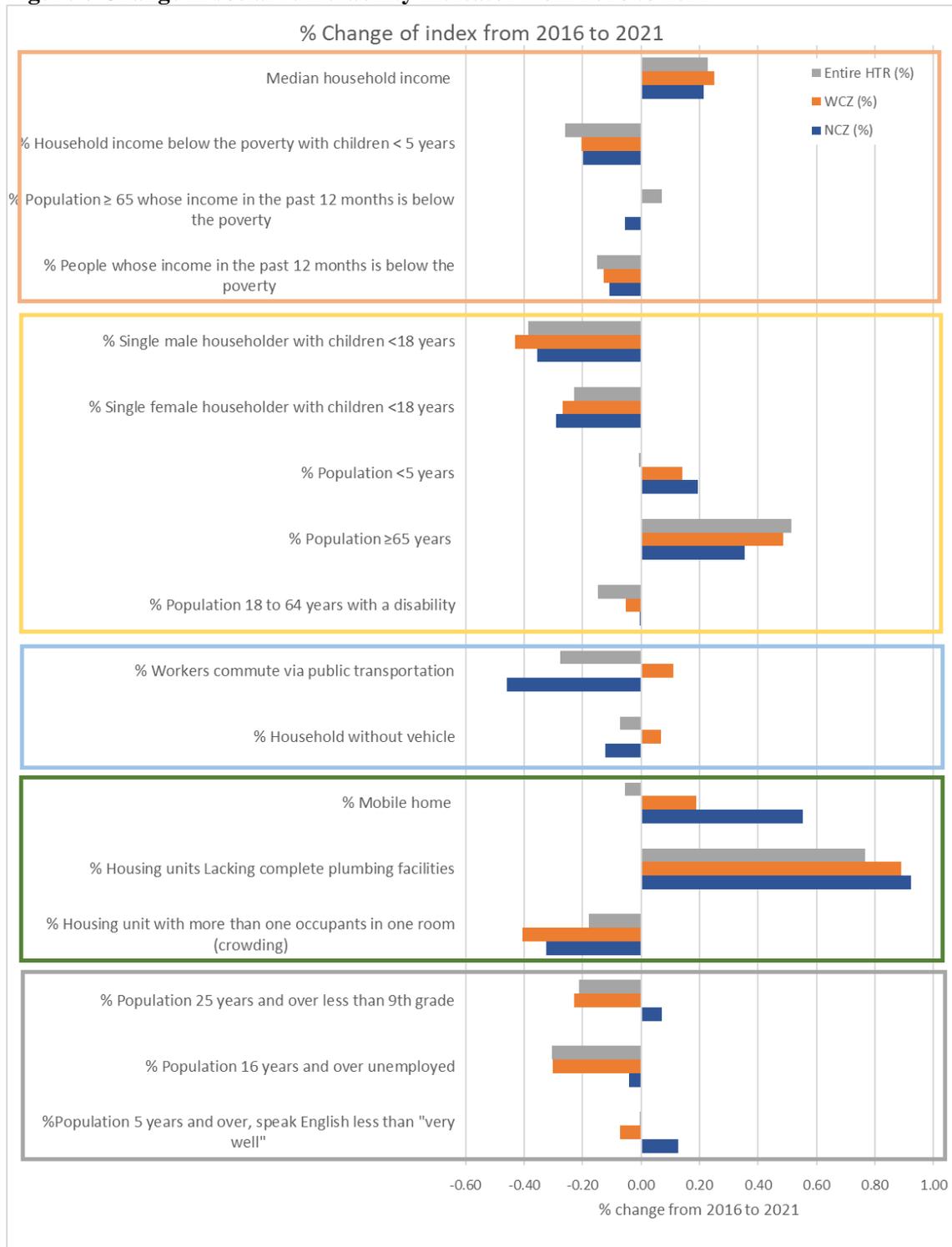

Table 4 shows social vulnerability characteristics of three geographical areas: the entire Hampton Roads region, the census tracts associated with WCZs, and the census tracts associated with NCZs. Each row contains the mean values of first order SVI input variables obtained from the 2016 and 2021 ACS data, and the rightmost columns show the percentage change between 2016 and 2021 for all SVI inputs. It is worth noting that while the overall Hampton Roads region experienced population growth between 2016 and 2021, the populations in census tracts associated with WCZs and NCZs both declined in the same 5-year period. The percentage change in median household income for all three geographic areas between 2016 and 2021 are relatively similar, increasing by 21% to 25%.

The blocks highlighted in gray in the rightmost columns show categories of social vulnerability that has disproportionately increased (as compared to the Hampton Roads region as a whole) between 2016 and 2021 in the critical zones with high accessibility reduction. A negative value in the percentage change column suggests that the corresponding social vulnerability indicator is decreasing. Meanwhile, positive values in the change columns indicate that the population in the geographic area has become more vulnerable between 2016 and 2021. According to Table 4 and Figure 6, 6 social vulnerability indicators show a deterioration in 2021 compared to 2016 for both WCZ and NCZ, compared to the Hampton Roads region. These 6 indices are as follows: percentage of household income below poverty with children less than 5 years old, percentage of people whose income in the past 12 months is below poverty, percentage of population less than 5 years old, percentage of 18- to 64-year-olds with disability, percentage of population residing in mobile homes, and percentage population 16 years and over who are unemployed. Previous studies by Erman et al. ( 2019) and Hallegatte et al. (2020) have indicated people vulnerable in socio-economic status are more likely to be affected with higher frequency of flooding, as they have limited ability to relocate from the repeated flood risk compared to their wealthier counterparts. These results might imply that individuals with such social vulnerability have limited options for relocating away from the critical zones, thereby becoming increasingly vulnerable to recurrent flooding impacts.

## 6. CONCLUSIONS AND LIMITATIONS

Existing literature overlooks the temporal effects of recurrent flooding on transportation systems and populations. This study used WAZE flood incident and congestion data from five Augusts between 2018 and 2022 to identify recurrent flooding "hotspot" locations in the Hampton Roads region, and to examine social vulnerability change in these highly impacted areas between 2016 and 2021.

The study yields several key findings. Firstly, areas with significant recurrent flooding-induced transportation accessibility reduction are predominantly within the city of Norfolk. Overall, it was found that approximately 11.95% and 3.09% of the total Hampton Roads population experience frequent reduction in transportation accessibility for work and non-work trips, respectively. Secondly, an analysis of the social vulnerability in critical recurrent flood-impacted areas (WCZs and NCZs) with census data from 2016 and 2021 highlights that socio-economically disadvantaged groups with limited transportation options are more likely to experience frequent and significant accessibility reduction. Lastly, comparing the change in social vulnerability for the populations living in WCZs and NCZs between 2016 and 2021, study results indicate that these populations are, on average, increasingly more socially vulnerable than the overall Hampton Roads region. The increased vulnerability in six indices may signify a reduced capacity for the population living in poverty and with mobility constraints to cope with the recurrent flooding events, due to a lack of relocation options.

This study offers novel insights into recurrent flooding impacts on both transportation infrastructure and population social vulnerability at specific regional time scales. By enhancing crowdsourced data processing workflow with customized ArcGIS Pro toolboxes, the study efficiently identifies areas most frequently experiencing significant accessibility reduction over multiple years. Furthermore, the study examines the temporal component of social vulnerability of populations in critical zones, highlighting trends in vulnerability indices and emphasizing the need to consider the dynamic nature of populations affected by climate events. Specifically, in the Hampton Roads region, the study reveals that the impacts of recurrent flooding are disproportionately experienced by those with low-income and limited mobility, a situation that may be deteriorating over time. These findings bear substantial implications for shaping future climate change mitigation strategies and recurrent flooding resilience management.

This case study has several limitations. First, the population data used in the accessibility calculation for TAZs are from 2015, thus any changes in population at the TAZ level between 2016 and 2021 are not accounted for in the accessibility reduction calculations. Note that ACS data for 2016 and 2021 are used in the SVI calculations, so changes in population across 5 years are accounted for in the SVI analysis. Second, the spatial unit of the ACS census data and HRTPO TAZ data are not the same, and the incongruity between TAZ and census tracts impairs the precision of the social vulnerability analysis. Using TAZ as base spatial unit for accessibility analysis is effective for recognizing the locations that are highly impacted by recurrent flooding in the transportation network, however this spatial unit might be too granular when examining location-specific community demographics. An enhancement to the current study might encompass broader geographic coverage with WAZE data, adopting census tracts as the base spatial unit. Given that the Hampton Roads region contains a relatively limited number of census tracts (i.e., 413), the sample size may be considered insufficient if only this region is included in the analysis. Moreover, the temporal scope (5 years) in this study is relatively limited. To understand how social vulnerability of a sub-population is evolving over time, a longer period of data collection would be needed in order to arrive at more robust conclusions. Lastly, the users of WAZE are not representative of the Hampton Roads region's overall population. Socially vulnerable population have limited access to smartphones (*39*, *40*), therefore flood incidents in certain areas can be underreported, and the actual impacts of flooding might be worse for socially vulnerable groups compare to what is analyzed in this study. This limitation also highlights a potential avenue for future research, a continued standardized collection of data and analysis on the impact of recurrent flooding on the transportation system and social vulnerability, providing continually updated results relevant to localized climate event management and mitigation policies.

**ACKNOWLEDGEMENTS**
The authors thank the Hampton Roads Transportation Planning Organization and WAZE for facilitation of data acquisition. Then authors also like to thank Yiqing Xu for giving valuable suggestions on ArcGIS pro model builder function. This work is supported by the National Science Foundation's Critical Resilient Interdependent Infrastructure Systems and Processes program (Award 1735587).

**AUTHOR CONTRIBUTIONS**
The authors confirm contribution to the paper as follows: study conception and design: L. Zeng, T.D. Chen, J. Miller, J.L. Goodall; data collection: F.T. Zahura; analysis and interpretation of results: L. Zeng, T.D. Chen; draft manuscript preparation: L. Zeng, T.D. Chen. All authors reviewed the results and approved the final version of the manuscript.